\documentclass[aps,prd,onecolumn,groupedaddress,preprintnumbers,superscriptaddress]{revtex4-2}  
\usepackage{graphicx,mathtools,tabu,array}
\usepackage{epstopdf}
\usepackage{amsmath}
\usepackage{amsfonts}
\usepackage[colorlinks=true,citecolor=red,linkcolor=blue,breaklinks=true]{hyperref}
\usepackage{accents}
\usepackage[figure, table]{hypcap}
\usepackage{amssymb}
\usepackage{appendix}
\usepackage{comment}
\usepackage{bbold}
\usepackage{xcolor}
\usepackage{slashed}
\usepackage{subfigure}
\usepackage{setspace}
\usepackage{footnote}
\usepackage{multirow}
\usepackage{braket}
\usepackage[normalem]{ulem}
\usepackage[utf8]{inputenc}
\usepackage{mathrsfs}
\usepackage{longtable}
\usepackage{enumitem}
\usepackage{orcidlink}
\usepackage{booktabs} 
\usepackage{subfigure}

\newcommand{\PreserveBackslash}[1]{\let\temp=\\#1\let\\=\temp}
\newcolumntype{C}[1]{>{\PreserveBackslash\centering}p{#1}}
\newcolumntype{R}[1]{>{\PreserveBackslash\raggedleft}p{#1}}
\newcolumntype{L}[1]{>{\PreserveBackslash\raggedright}p{#1}}

\LTcapwidth=\textwidth

\allowdisplaybreaks

\definecolor{palatd}{RGB}{104, 36, 109}
\definecolor{palatb}{RGB}{0, 56, 168}
\definecolor{palatr}{rgb}{0.745,0.118,0.176}
\newcommand\myshade{80}
\colorlet{mylinkcolor}{palatr}
\colorlet{mycitecolor}{palatb}
\colorlet{myurlcolor}{palatd}

\hypersetup{
  linkcolor  = mylinkcolor!\myshade!black,
  citecolor  = mycitecolor!\myshade!black,
  urlcolor   = myurlcolor!\myshade!black,
  colorlinks = true
}

\begin{document}
\sloppy  

\preprint{FT-UAM/CSIC-26-71}

\title{Bose-enhanced Neutrino Decays in a Thermal Medium}

\author{Yuber F. Perez-Gonzalez\,\orcidlink{0000-0002-2020-7223}}
\email{yuber.perez@uam.es}
\affiliation{Departamento de F\'{i}sica Te\'{o}rica and Instituto de F\'{i}sica Te\'{o}rica (IFT) UAM/CSIC, Universidad Aut\'{o}noma de Madrid, Cantoblanco, 28049 Madrid, Spain}

\author{Manibrata Sen\,\orcidlink{0000-0001-7948-4332}}
\email{manibrata@iitb.ac.in}
\affiliation{Department of Physics, Indian Institute of Technology Bombay, Powai, Maharashtra 400076 India}

\author{Walter~Tangarife\,\orcidlink{0000-0002-4808-3277}}
\email{wtangarife@luc.edu}
\affiliation{Department of Physics, Loyola University Chicago, Chicago, IL 60660, USA}

\begin{abstract}
We compute the decay width of neutrinos in a thermal medium using finite-temperature quantum field theory, focusing on non-standard decays into lighter neutrinos and a scalar or light vector boson. We derive general expressions for the thermal decay rate and show that finite-temperature effects can dramatically enhance neutrino decays when the parent and daughter states are nearly degenerate in mass. In this regime, the emitted boson is kinematically soft and undergoes strong Bose enhancement, leading to decay widths that can exceed their vacuum values by orders of magnitude. We demonstrate that this effect is largely insensitive to the Lorentz structure of the underlying interaction and arises generically from the interplay of thermal occupation factors and quasi-degenerate kinematics. Our results highlight a previously underappreciated feature of neutrino decay in thermal environments and provide a general framework applicable to a broad class of fermionic decay processes.

\end{abstract}

\maketitle

\section{Introduction}
\label{sec:intro}

Neutrinos provide a crucial clue in our quest for physics beyond the Standard Model (SM). The discovery of neutrino oscillations, and hence non-zero neutrino mass, points to the presence of new physics beyond our current understanding. The existence of neutrino mass implies that they can decay. In the framework of three active neutrinos, the heavier mass 
eigenstates can decay into lighter ones, with the decay channels depending 
on the neutrino mass ordering. In the normal ordering, the state $\nu_3$ 
(and potentially $\nu_2$) can decay into lighter states, while in the 
inverted ordering the state $\nu_2$ (and $\nu_1$) can undergo similar 
transitions.

Within SM interactions, neutrino decays proceed through a one-loop diagram and hence, neutrino lifetimes can be very large, comparable to, or larger than the age of the Universe~\cite{Pal:1981rm}. These decays can either involve radiative processes such as $\nu_i \to \nu_j \gamma$, 
or non-radiative ones with two to three neutrinos in the final 
state. Non-radiative decays are usually
classified as either invisible or visible depending on whether the decay 
products can be detected. Invisible decays typically manifest as a depletion 
of the expected neutrino flux, while visible decays modify the observed 
energy and flavor distributions in detectors. However, in many extensions of the SM, neutrinos may decay into lighter states through interactions with new bosons and hence can lead to a faster decay~\cite{Schechter:1981cv,PhysRevLett.45.1926,GELMINI1981411,Gelmini:1983ea,BERTOLINI1988714,SANTAMARIA1987423}. This makes non-standard decays of neutrinos one of the most fascinating probes of new physics in the neutrino sector. 

Neutrino decay has been extensively studied in both laboratory and astrophysical settings. Existing constraints have been derived from solar, atmospheric, reactor, accelerator, supernova, and cosmological observations~\cite{1984MNRAS.211..277D, Doroshkevich:1989bf, Berezhiani1992, Fogli:1999qt,Choubey:2000an,Lindner:2001fx,Beacom:2002cb,Joshipura:2002fb,Bandyopadhyay:2002qg, Beacom:2002vi,Beacom:2004yd,Berryman:2014qha,Picoreti:2015ika,FRIEMAN1988115,Mirizzi:2007jd,GonzalezGarcia:2008ru,Maltoni:2008jr,Baerwald:2012kc,Broggini:2012df,Dorame:2013lka,Gomes:2014yua,Abrahao:2015rba,Coloma:2017zpg,Gago:2017zzy,Choubey:2018cfz,Ascencio-Sosa:2018lbk,Chianese:2018luo,deSalas:2018kri,Escudero:2019gfk, deGouvea:2019goq,Funcke:2019grs,Escudero:2020ped,Abdullahi:2020rge,Akita:2021hqn,Barenboim:2020vrr, Picoreti:2021yct,DeGouvea:2020ang, Chen:2022idm,Ivanez-Ballesteros:2022szu, deGouvea:2022cmo,deGouvea:2023jxn, Roux:2024zsv,Telalovic:2024cot}. These studies typically constrain invisible or visible decay channels by comparing the observed neutrino fluxes and spectra with expectations from the standard no-decay hypothesis. The absence of significant deviations from predictions translates into stringent bounds on neutrino lifetimes and on the underlying interactions responsible for the decay. 

A common feature of most existing studies of neutrino decay is the assumption that the decay occurs in vacuum. However, inside compact astrophysical objects, or in the early Universe, neutrinos propagate in a thermal background consisting of relic neutrinos and other relativistic species~\cite{PhysRevD.17.2369,Mikheev:1986gs,Duan:2006an,Hannestad:2006nj,Mirizzi:2015eza, Volpe:2023met, Sen:2024fxa,Farzan:2002wx,Ivanez-Ballesteros:2025ojj,Raffelt:2025wty}. This can also be relevant in scenarios beyond the SM, where hidden sector plasmas may arise. In this environment, decay processes are modified by statistical effects arising from the particle occupation numbers in the thermal bath. These effects include Pauli blocking, which suppresses processes that produce fermions in already occupied states, and Bose enhancement, which can increase decay rates when bosonic final states are present in the thermal bath. These phenomena are naturally described within the framework of finite-temperature quantum field theory, where interaction rates are determined by the imaginary part of the particle self-energy.

Thermal effects may become particularly important for neutrino decay in the quasi-degenerate regime where the mass difference between neutrino eigenstates is small. In this limit, the available phase space for the decay is suppressed in vacuum, making the decay rate sensitive to even modest modifications from the surrounding thermal environment. The softness of the resulting bosonic particles also adds to a Bose enhancement of the decay rate, and this can significantly exceed the vacuum contribution. Understanding these corrections is therefore necessary for the accurate interpretation of cosmological bounds on non-standard neutrino interactions. 

Medium effects on neutrino decay have been studied before in the context of the radiative channel $\nu_i \to \nu_j\gamma$. Although this decay is suppressed in vacuum, via the GIM mechanism, coherent interaction of neutrinos with a charged-lepton plasma induces an effective electromagnetic vertex that opens the channel and enhances the rate relative to its vacuum value~\cite{DOlivo:1989brs}. The
stimulated emission of soft photons from the thermal bath, leading to additional Bose enhancement of the decay rate, was studied in~\cite{Nieves:1997md,Grasso:1998td}.

In this work, we revisit fermion non-radiative decay in a thermal background, in the presence of beyond-SM interactions, and investigate the impact of thermal effects on non-standard neutrino decay processes. We derive the decay rate using finite-temperature field theory by relating the imaginary part of the fermion self-energy to the interaction rate in a thermal bath. This formalism allows us to obtain a general expression for the decay rate that includes both the standard vacuum contribution and additional terms arising from thermal occupation numbers. We then apply the resulting expression to neutrino decays mediated by both scalar and vector interactions. We find that in the limit where the parent and daughter neutrinos are quasi-degenerate, the thermal decay rates are indeed enhanced and can modify the decay phenomenology.  We show that this enhancement is largely independent of the detailed nature of the bosonic mediator and arises as a generic consequence of finite-temperature effects combined with quasi-degenerate kinematics.
This may have important consequences for neutrino decays in the early Universe or inside core-collapse supernovae. 

The formalism developed here is not restricted to neutrino physics. More generally, particle decay in a thermal background occurs in many contexts across early-Universe cosmology and particle physics. For example, similar effects can arise in models of multi-component dark matter where a heavier dark-sector particle decays into lighter species in the presence of a thermal bath. In such scenarios, thermal corrections may modify decay rates and thereby influence the cosmological evolution of the dark sector. The framework presented in this work can therefore be readily adapted to a wide class of new-physics scenarios involving particle decay in a thermal environment.

The paper is organized as follows.
We present the derivation of the thermal decay width for neutrino decay into a lighter one and a boson using finite-temperature field theory in Sec.~\ref{sec:formalism}.
We describe in detail the computation for a scalar in the final state in Secs.~\ref{subsec:scalar},~\ref{subsec:thermalmass}, and provide results for the vector case in Sec.~\ref{subsec:vector}.
A few appendices are included. In Apps.~\ref{app:calculation} and~\ref{app:vector}, we give details of the computation performed here; and in App.~\ref{app:enhancement}, we elaborate on the physical origin of the enhancement found for the thermal decay width.

\section{Fermion Decay in a Thermal Background}
\label{sec:formalism}
The decay rate of a particle is related to the imaginary part of its self-energy. For a fermion with mass $M$, the full propagator can be written as
\begin{equation}
  S^{-1}(p) = \slashed{p} - M - \Sigma(p),
\end{equation}
where $\Sigma(p)$ is the self-energy matrix. The decay rate is obtained from the imaginary part of the projected self-energy. The physical decay rate arises from the imaginary part of the self-energy. In particular, the optical theorem implies that the imaginary part of the on-shell self-energy determines the total interaction rate of the particle. The quantity relevant for the decay rate is the self-energy~\cite{Weldon:1983jn}
\begin{equation}
\Pi(p) = \frac{1}{2}\sum_{\rm spin} \overline{u}(p) \,\Sigma (p) \,u(p) 
= \frac{1}{2}{\rm Tr}\left[(\slashed{p}+M) \Sigma(p)\right]\,.
\end{equation}
The decay rate $\Gamma_D$ is related to the imaginary part of this quantity through
\begin{equation} \label{eq:dacay-def}
\Gamma_D(\omega)\,=\,-\frac{1}{1+e^{-\omega/T}}  \frac{\operatorname{Im} \Pi(\omega)}{\omega},
\end{equation} where $\omega$ is the energy. In general, $\operatorname{Im}\Pi(\omega)$ provides both direct and inverse decay processes, which can be related by $\Gamma_I\,=\,e^{-\omega/T}\,\Gamma_D$, where $\Gamma_I$ is the inverse decay rate. For fermions in a thermal bath, the sum of the rates, $\Gamma_D+\Gamma_I$, provides the rate at which the particle's population relaxes toward equilibrium~\cite{Weldon:1983jn}. We refer the reader to Appendix~\ref{app:calculation} for more details. 

The imaginary part can be obtained by calculating the discontinuity of the self-energy~\cite{Bellac:2011kqa}, 
\begin{equation} \label{eq:discontinuity}
    \operatorname{Disc}\Pi(\omega)\,=\,2i \operatorname{Im}\Pi(\omega),
\end{equation}

Therefore, calculating the decay rate requires calculating the self-energy matrix $\Sigma(p)$. We now apply the general formalism to neutrino decays in extensions of the SM. We consider scenarios in which a heavier neutrino mass eigenstate decays into a lighter neutrino and a new bosonic particle. 

\subsection{Neutrino decays into a light scalar boson}\label{subsec:scalar}
The relevant Lagrangian for neutrino decays into a light scalar boson can be written in the form of a Yukawa-like interaction as  
\begin{equation} \label{eq:Lagrangian-int}
    \mathcal{L}=g_{ij}\,\overline{\nu_{iL}}\nu_{jR} \phi\,,
\end{equation} where $g_{ij}$ is a dimensionless matrix of couplings that connects the different mass eigenstates $\nu_{i}$, $i=\{1,2,3\}$.
For simplicity, we assume from now on that all the couplings are equal $g\equiv |g_{ij}|$.
Another relevant operator leading to similar decays is given by $\mathcal{L}=g_{ij}\,\overline{\nu_{iL}^c}\nu_{jL} \phi\,$. Such interactions appear in Majoron models and other frameworks with light scalar mediators.

\begin{figure}[!t]
\centering
\includegraphics[width=0.5\textwidth]{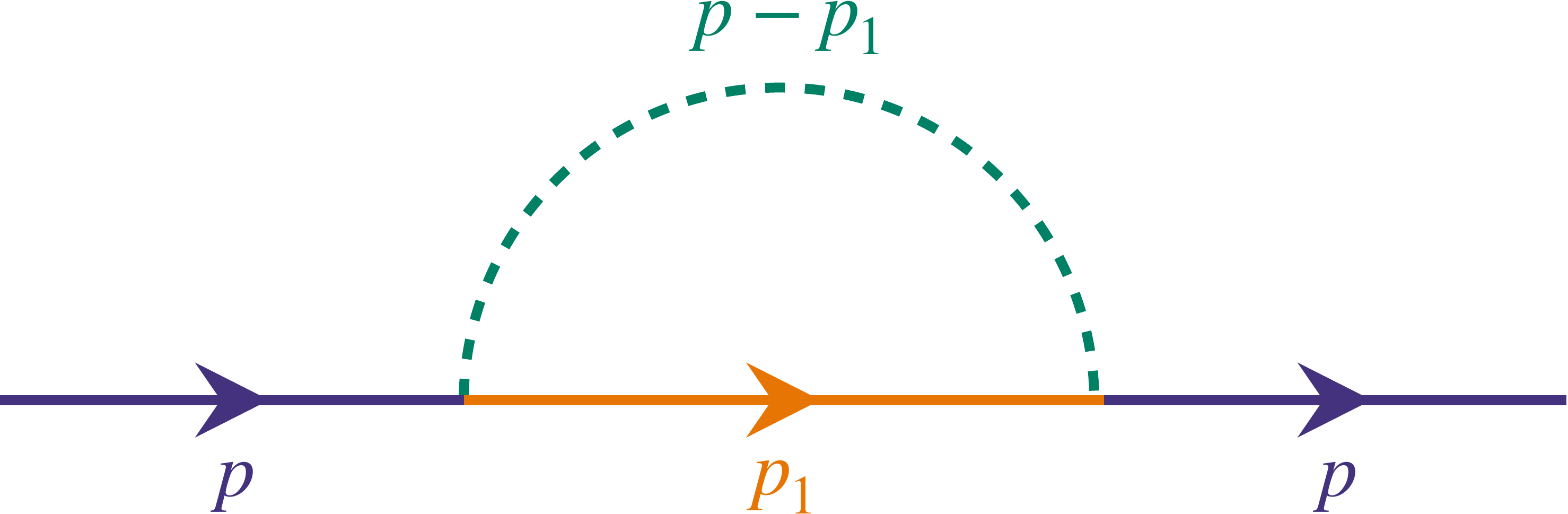} 
\caption{One-loop contribution to the neutrino self-energy. The intermediate particle in the loop can either be a scalar or a vector boson.}
\label{fig:self-energy}
\end{figure}
The self-energy contribution comes from the neutrino's interaction with a scalar field, as shown in Fig.~\ref{fig:self-energy}, and can be written as
\begin{equation}
\label{eq:self-energy-def}
    -i \Sigma(p)\,=\,g^2 \int \frac{dp^0_1}{2\pi} \int \frac{d^3p_1}{(2\pi)^3}\,\mathrm{P_R }\,S(p_1)\,\mathrm{P_L }\,D(p-p_1), 
\end{equation}
where $S(p)$ and $D(p)$ are the fermionic and scalar propagators respectively. To evaluate these kind of diagrams, we use the imaginary-time formalism in which frequencies are analytically extended to imaginary discrete values, $p^0\,=\,i \omega_n\,=\,2n\beta \pi\,i $ for bosons, and $p^0\,=\,i \omega_n\,=(2n+ 1)\beta \pi\,i$ for fermions, also known as Matsubara frequencies\,\footnote{We have also confirmed that our results agree with those derived in the real-time formalism.}.

Implementing the Matsubara formalism, these propagators become~\cite{Bellac:2011kqa}
\begin{eqnarray}
 S(p_1)&=&\frac{\slashed{p}+m}{p_1^2-m^2}  \,=\,- \frac{i\omega_n\gamma^0 - \vec{p_1}\cdot \vec{\gamma} +m}{\omega_n^2+ E_1^2} \,, \quad {\rm  and}\\
D(p-p_1) &=& \frac{1}{(p-p_1)^2-m_\phi^2}\,=\,-\frac{1}{(\omega-\omega_n)^2+E_2^2}\,,
\end{eqnarray}
where $E_1^2\equiv \vec{p}_1^{\,2}+m^2$, and $E_2^2\,\equiv\, (\vec{p}-\vec{p}_1)^{2}+m_\phi^2$. 

The integral over $p^0_1$ in Eq.~\eqref{eq:self-energy-def} is carried out as a sum using the identities~\cite{Bellac:2011kqa}
\begin{footnotesize}
\begin{eqnarray}
    T\sum_{n}^{\infty} \frac{1}{\omega_n^2+E_1^2} \frac{1}{(\omega-\omega_n)^2+E_2^2} &=& \frac{1}{4E_1 E_2}\left\lbrace \left(\frac{1}{i\omega +E_1-E_2} - \frac{1}{i\omega -E_1+E_2}\right)\left(f_{FD}(E_1)+f_{BE}(E_2) \right) \right. \nonumber\\
    && + \left.\left(\frac{1}{i\omega +E_1+E_2} - \frac{1}{i\omega -E_1-E_2}\right)\left(1-f_{FD}(E_1)+f_{BE}(E_2) \right)  \right\rbrace,  \\
    &\equiv& F_1(i\omega, E_1,E_2) \nonumber\\
    T\sum_{n}^{\infty} \frac{\omega_n}{\omega_n^2+E_1^2} \frac{1}{(\omega-\omega_n)^2+E_2^2} &=& \frac{i}{4E_2}\left\lbrace \left(\frac{1}{i\omega +E_1-E_2} + \frac{1}{i\omega -E_1+E_2}\right)\left(f_{FD}(E_1)+f_{BE}(E_2) \right) \right. \nonumber\\
    && + \left.\left(\frac{1}{i\omega +E_1+E_2} + \frac{1}{i\omega -E_1-E_2}\right)\left(1-f_{FD}(E_1)+f_{BE}(E_2) \right)  \right\rbrace \\
        &\equiv& i\,F_2(i\omega, E_1,E_2) \nonumber
 \end{eqnarray}
\end{footnotesize}
where $f_{FD}$ and $f_{BE}$ are the Fermi-Dirac and Bose-Einstein distribution functions respectively. In our calculations, we assume the chemical potential of the neutrinos to be zero, corresponding to a vanishing particle-antiparticle asymmetry. Note that a non-zero chemical potential can be incorporated straightforwardly by replacing the equilibrium distribution functions with their generalized forms. This modifies the occupation numbers quantitatively but does not alter the conclusions qualitatively.

After the sum, we obtain
\begin{equation} \label{eq:self-ene-3}
\Sigma(p)\,=-\,g^2  \int \frac{d^3p_1}{(2\pi)^3} (-\gamma^0 F_2(i\omega, E_1,E_2) - \vec{p_1}\cdot \vec{\gamma} F_1(i\omega, E_1,E_2)) \mathrm{P_L}.
\end{equation}

Using the results above and Eq.~\eqref{eq:dacay-def}, the net decay rate of a neutrino of average energy, $E$, in a thermal bath evaluates to 
\begin{eqnarray}
    \Gamma_D(E)&=&  \frac{g^2 (M^2+m^2-m_\phi^2)}{32\pi\,E\,\sqrt{E^2-M^2}}  \frac{T}{1+e^{-E/T}} \log \left[\frac{\left(1+e^{\omega_{+}/T}\right)\left(e^{E/T}-e^{\omega_{-}/T}\right)}{\left(1+e^{\omega_{-}/T}\right)\left(e^{E/T}-e^{\omega_{+}/T}\right)} \right],
 \label{eq:GammD}
\end{eqnarray}
where 
\begin{equation}
 \omega_{\pm} \,=\, \frac{E}{2}\left(1-\frac{m_\phi^2-m^2}{M^2}\right)\pm \frac{\sqrt{E^2-M^2}}{2}\sqrt{\left(1-\frac{(m_\phi-m)^2}{M^2}\right)\left(1-\frac{(m_\phi+m)^2}{M^2}\right)}\, ,
 \label{eq:wpm}
\end{equation}
and $M$ denotes the mass of the heaviest neutrino. 

\begin{figure}
    \centering
    \includegraphics[width=0.9\linewidth]{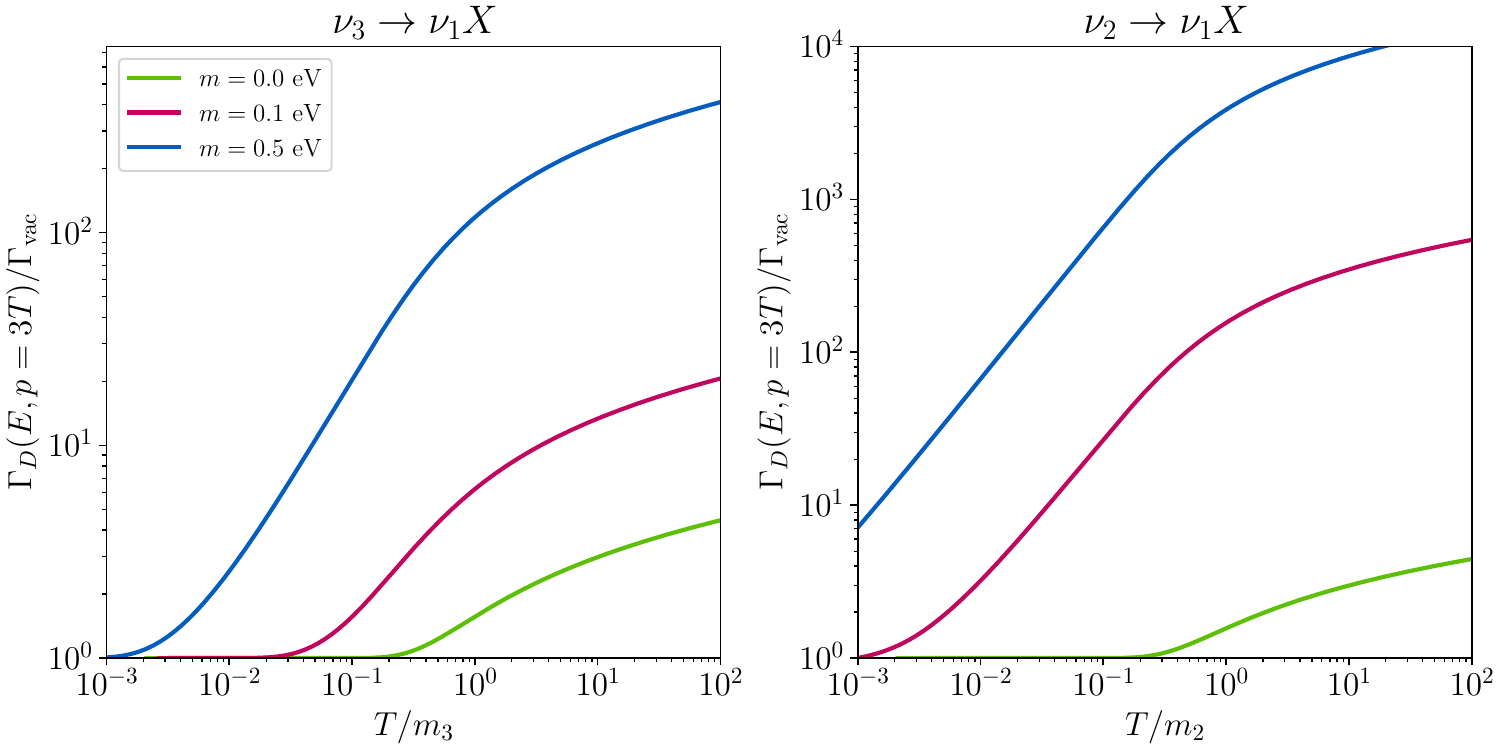}
    \caption{Ratio between the thermal and vacuum decay rates for $\nu_3\to\nu_1X$ (left) and $\nu_2\to\nu_1X$ (right) as a function of the temperature normalized to the parent neutrino mass, assuming the lightest neutrino mass to be  $m = 0$ (green) $m = 0.1~{\rm eV}$ (dark red), $m=0.5~{\rm eV}$ (blue), and $m_\phi=0$.}
    \label{fig:ratio_neutrino}
\end{figure}

We present in Fig.~\ref{fig:ratio_neutrino} the ratio of the net decay rate to the vacuum value $\Gamma_{\rm vac}$,
\begin{eqnarray}
    \Gamma_{\rm vac}(E)&=&  \frac{g^2 (M^2+m ^2-m_\phi^2)}{32\pi\,E}  \sqrt{\left(1-\frac{(m-m_\phi)^2}{M^2}\right)\left(1-\frac{(m+m_\phi)^2}{M^2}\right)} ,
 \label{eq:Gamm_vac}
\end{eqnarray}
as function of the plasma temperature $T$ normalized to the parent neutrino mass $m_i$, $i=\{3,2\}$. Neutrino masses are fixed from neutrino oscillations and the lightest neutrino with mass $m$. We have taken the quadratic mass splittings to be $\Delta m^2_{21} = 7.5\times 10^{-5}~{\rm eV^2}$ and $\Delta m^2_{31} = 2.5\times 10^{-3}~{\rm eV^2}$, following the recent results from the NuFIT group~\cite{Esteban:2024eli}. In the left panel of Fig.~\ref{fig:ratio_neutrino}, we present the ratio for the $\nu_3\to\nu_1 X$ decay, while in the right panel we show the $\nu_2\to\nu_1 X$ case. Here, $X$ represents either the scalar considered in this section or a light vector that we analyze in the next section.
We take the momentum of the parent neutrino to be $p=3T$ as a representative momentum for a particle in thermal equilibrium. We have verified that averaging this ratio with a Fermi-Dirac distribution does not significantly alter our results.
Note that for $\nu_2\to\nu_1 X$ decay, the mass ratio can be as small as $m_2/m_1 = 1.0001$  for $m_1=0.5~{\rm eV}$, while for $\nu_3\to\nu_1 X$ decay, we have $m_3/m_1 = 1.005$. These ratios illustrate that the neutrino spectrum becomes highly quasi-degenerate when the absolute neutrino mass scale approaches the eV range. For temperatures larger than the parent neutrino mass, $T\gtrsim m_i$, we find a significant enhancement of the decay width relative to its vacuum value. For the most degenerate spectra considered here, this enhancement reaches values of several orders of magnitude $\Gamma_D/\Gamma_{\rm vac}\gtrsim \mathcal{O}(10^3)\,{\rm or}\,\mathcal{O}(10^3)$ for $\nu_3\to\nu_1 X$ and $\nu_2\to\nu_1 X$, respectively. In general, the ratio plateaus for very high temperatures, $T\gtrsim 10^3 \,m_i$.
%
\begin{figure}
    \centering
    \includegraphics[width=0.9\linewidth]{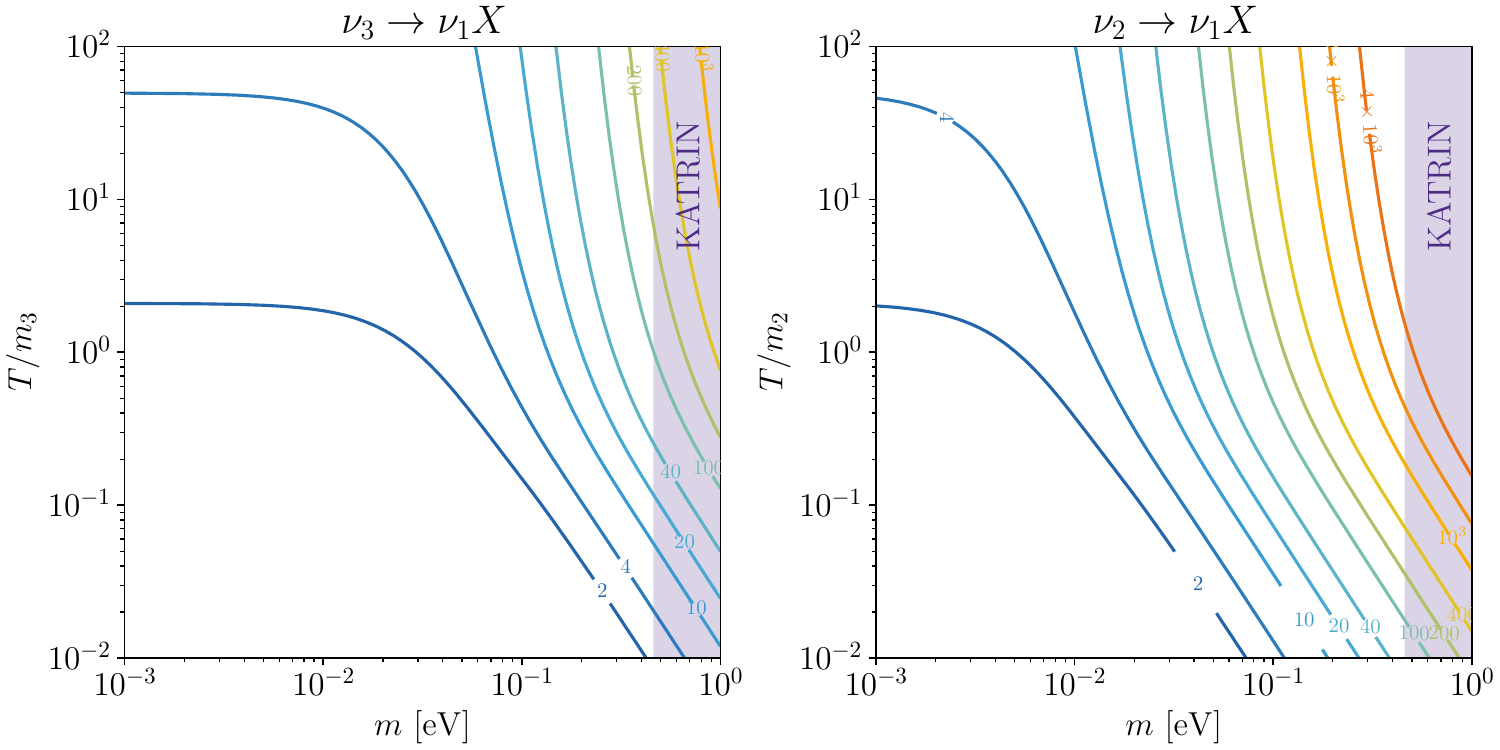}
    \caption{Ratio of the thermal to vacuum decay rates for $\nu_3\to\nu_1X$ (left) and $\nu_2\to\nu_1X$ (right) as a function of temperature normalized to the parent neutrino mass and the lightest neutrino mass $m$. We have assumed $m_\phi=0$. The purple region corresponds to values currently excluded by the KATRIN experiment~\cite{KATRIN:2024cdt}. }
    \label{fig:contour_ratio}
\end{figure}
Fig.~\ref {fig:contour_ratio} shows the enhancement as a function of plasma temperature normalized to the parent neutrino mass and the lightest neutrino mass. The purple band indicates the values currently excluded by the KATRIN experiment~\cite{KATRIN:2024cdt}.  The plot illustrates how the constraints still allow enhancements of several orders of magnitude for decays of $\nu_2$ and $\nu_3$.

We now briefly summarize the physical origin of the thermal enhancement appearing in the decay rate (for details, see Appendix \ref{app:enhancement}). In the rest frame of the parent neutrino, energy-momentum conservation fixes the momentum of the final state particles. For a light scalar and in the quasi-degenerate limit, defined by a small mass splitting $\delta = M - m \ll M$, the momentum of the emitted scalar is suppressed and scales as $p \sim \delta$. As a result, the scalar is kinematically soft in the rest frame of the parent particle.  Boosting to the frame where the parent neutrino is moving with an energy $E$, the energy of the scalar depends on the emission angle relative to the parent 
momentum. In particular, for backward emission, one finds that the scalar energy is further suppressed by the Lorentz factor, leading to 
\begin{equation}
    E_\phi \simeq\frac{\delta}{\gamma}\,,
\end{equation}
where $\gamma = E/M$ is the boost factor of the parent neutrino. For relativistic neutrinos, $\gamma \gg 1$, implying that the scalar can become extremely soft in the lab frame. Notice that this rest-frame argument is used here only to build intuition about the source of enhancement in the phase space. In the rest frame of the parent neutrino, the thermal bath is not isotropic, which introduces preferred boosts in the distribution functions and a direction-dependent effective temperature. Nevertheless, the arguments about the softness of the emitted boson still hold in the plasma reference frame, in which we performed all our calculations.

This kinematic suppression plays a crucial role in the thermal decay rate. The finite-temperature correction arises through the statistical factor appearing in the decay rate, $(1-f_{FD}+f_{BE})$. When the scalar energy satisfies $E_\phi \ll T$, the BE distribution becomes large,
\begin{equation}
f_B(E_\phi) \simeq \frac{T}{E_\phi} \gg 1,
\label{eq:enhancement}
\end{equation}
leading to a significant enhancement of the decay rate.

\begin{figure}[!t]
\centering
\includegraphics[width=0.5\textwidth]{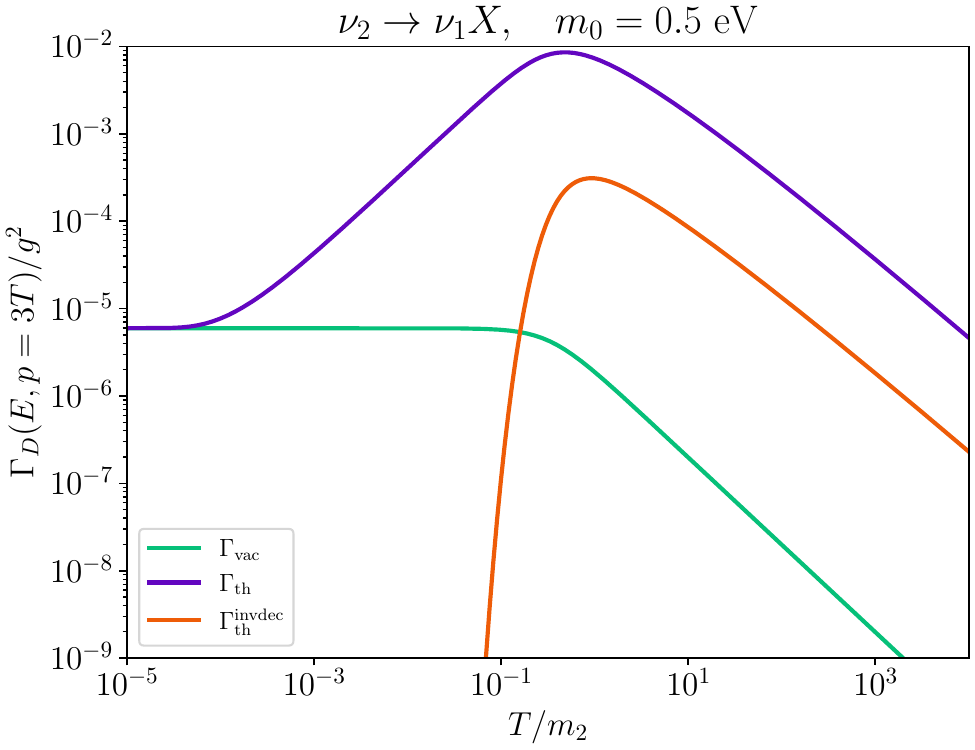} 
\caption{ Vacuum (green), thermal (purple), and thermal inverse (orange) decay rates for $\nu_2\leftrightarrow\nu_1X$, as a function of the temperature normalized to the parent neutrino mass, assuming the lightest neutrino mass to be $m=0.5~{\rm eV}$}
\label{fig:abs-rates}
\end{figure}

The logarithmic terms in Eq.\,(\ref{eq:GammD}) can be simplified in this limit by expanding the thermal factors for 
small energy transfer. In the quasi-degenerate regime, the relevant kinematic combinations entering the logarithm are proportional to $\delta$, resulting in an additional enhancement. For a relativistic parent neutrino, $E \gtrsim T$, which leads to an approximate scaling,
\begin{equation}
\frac{\Gamma_{D}}{\Gamma_{\rm vac}} \simeq \frac{1}{2}e^{E/T}\,,
\end{equation}
indicating that the enhancement can become substantial when the parent particle is relativistic. It is worth noting that the decay rate in vacuum, Eq.~\eqref{eq:Gamm_vac}, is strongly suppressed in the quasi-degenerate limit and it would seem that the large enhancement of the ratio $\Gamma_D/\Gamma_{\rm vac}$ is due to such suppression. However, the thermal decay width does not vanish in that limit. Therefore, the large relative enhancement reflects the medium effect of removing the kinematic suppression, rather than just a large factor multiplying a negligible rate. To illustrate the behavior of the rates in this limit, we show in Fig.~\ref{fig:abs-rates} the absolute values of the vacuum and thermal decay rates for $\nu_2\to\nu_1X$. For comparison, we also show the inverse decay process, which is clearly suppressed relative to the thermal decay, as expected from the relation $ \Gamma_I \,=\,e^{-\omega/T}\,\Gamma_D$.
%
%


\subsection{Effect of thermal mass in the thermal decay rate }\label{subsec:thermalmass}

\begin{figure}[t]
    \centering
    \includegraphics[width=0.9\linewidth]{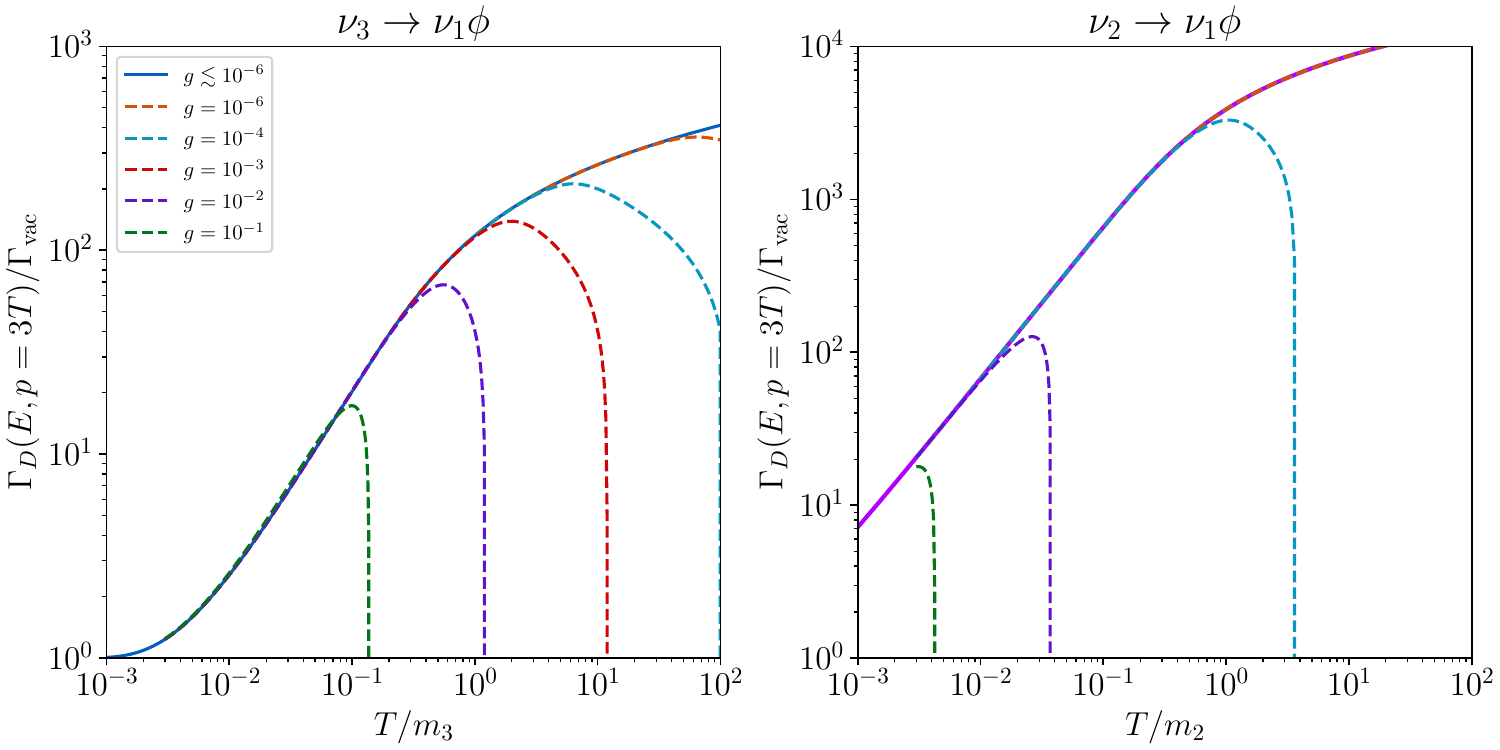}
    \caption{Effect of including thermal masses in the calculation of the thermal decay rate. The plot shows the ratio of the thermal (with thermal masses) to vacuum decay rates for $\nu_3\to\nu_1\,\phi$ (left) and $\nu_2\to\nu_1\,\phi$ (right) as a function of temperature. The effect is illustrated for $m=0.5\,\,{\rm eV}$, for which the decay rate enhancement is more prominent. }
    \label{fig:thermal_mass_ratio}
\end{figure}

The thermal enhancement discussed in Sec.~\ref{subsec:scalar} is modified for sufficiently large values of the coupling \(g\) due to thermal corrections to the masses of the bath fields~\cite{Bellac:2011kqa},
\begin{equation}
\label{eq:thermal_mass}
    \delta m^2(T)\simeq \frac{g^2}{16}T^2,
    \qquad
    \delta m_\phi^2(T)\simeq \frac{g^2}{6}T^2.
\end{equation}
These corrections arise from the real part of the finite-temperature self-energies $({\rm Re}\,\Sigma)$, and determine the thermal quasiparticle masses. Their inclusion, therefore, provides an improved description of the decay width in the thermal plasma, complementing the contribution of the imaginary part of the self-energy, which directly determines the decay rate. Although both effects originate from the same interaction and enter at the same perturbative order, the thermal mass corrections can have a substantial impact whenever the decay occurs close to a kinematic threshold, since even a modest shift in the masses may significantly alter the available phase space.

As shown in Fig.~\ref{fig:thermal_mass_ratio}, the effect of the thermal masses is negligible for small values of the coupling, where the thermal corrections remain much smaller than the vacuum masses. As $g$ increases, however, the thermal mass shifts become significant and progressively reduce the available phase space for the decay. Consequently, the thermal enhancement discussed above is progressively suppressed and may eventually disappear altogether once the threshold condition,
\begin{equation}
    m_{\nu_j}(T) > m_{\nu_1}(T) + m_X(T)\,,
\end{equation}
is not satisfied. In this regime, the decay becomes kinematically forbidden, and the thermal enhancement is effectively shut off.
Additionally, we observe that this suppression becomes significant for couplings $g \gtrsim 10^{-3}$. Whether this regime is phenomenologically relevant is model-dependent, as couplings of this size may be subject to experimental, cosmological, or theoretical constraints. Therefore, the impact of thermal-mass effects must be assessed within the context of a specific model realization.

\subsection{Neutrino decays into a light vector boson}\label{subsec:vector}

In the case of neutrinos coupled to a light vector, $A^\mu$, we replace the scalar propagator in Eq.~\eqref{eq:self-energy-def} by the gauge boson propagator,
\begin{equation}
 D_{\mu\nu}(q)\,=\, \frac{-\eta_{\mu\nu}+\frac{q_\mu q_\nu}{m_A^2}}{q^2-m_A^2}\,.
\end{equation}
This results in the net decay rate
\begin{equation} \label{eq:vector-mediator}
  \Gamma_D(E)\,=\,   \frac{g^2}{16\pi E \sqrt{E^2-m_A^2}}  \left(M^2+m^2-2m_A^2+\frac{(M^2-m^2)^2}{m_A^2}\right) \frac{T}{1+e^{-E/T}}\log \left[\frac{\left(1+e^{\omega_{+}/T}\right)\left(e^{E/T}-e^{\omega_{-}/T}\right)}{\left(1+e^{\omega_{-}/T}\right)\left(e^{E/T}-e^{\omega_{+}/T}\right)} \right], 
\end{equation} 
while the decay rate in vacuum is given by
\begin{eqnarray} \label{eq:vector-mediator-vac}
 \Gamma_{\rm vac}(E)&=& \frac{g^2}{16\pi E}  \left(M^2+m^2-2m_A^2+\frac{(M^2-m^2_\nu)^2}{m_A^2}\right) \sqrt{\left(1-\frac{(m-m_A)^2}{M^2}\right)\left(1-\frac{(m+m_A)^2}{M^2}\right)} \,,
\end{eqnarray} where $\omega_\pm$ are given by the expressions in Eq.~\eqref{eq:wpm}, after swapping $m_\phi$ with $m_A$.
Notice that the rate $\Gamma_D(E)/\Gamma_{\rm vac}$ coincides with the same function as in the scalar-mediator case, so the effects are the same as those shown in Figs.~\ref{fig:ratio_neutrino} and~\ref{fig:contour_ratio}.

The radiative decay of a neutrino into a vector boson in a medium was considered previously in Ref.~\cite{DOlivo:1989brs}. In that work, coherent forward scattering off the background induces an enhancement in the radiative channel, which is suppressed by the Glashow–Iliopoulos–Maiani (GIM) mechanism in vacuum. Furthermore, in Ref.~\cite{Nieves:1997md}, it was shown that a thermal bath of photons and electrons induces an enhancement of the decay rate due to the stimulated emission of soft photons, which was encoded in a factor $1+f_{BE}$ in the decay amplitude. Our result for the case of a light vector mediator shares similar features. In our case, the enhancement enters the amplitude as $1+f_{BE}-f_{FD}$ and the neutrino-vector interaction is a fundamental beyond-SM coupling rather than a medium-induced vertex. 

It is worth emphasizing here that the enhancement is largely independent of the underlying Lorentz structure of the interactions and is determined primarily by the thermal occupation numbers of the particles in the thermal bath. This observation suggests that the effect studied in this paper should be relevant for a broad class of non-standard fermion decays in a thermal environment.

\section{Discussion and conclusions}

Neutrinos are expected to decay even in the Standard Model. If physics beyond-the-Standard Model does not affect neutrino lifetime, we do not need to worry about decays, as such lifetimes are much larger than the age of the Universe.
However, additional interactions that neutrinos might exhibit can significantly alter such lifetimes. While most of the previous literature deals with non-standard decays of neutrinos in vacuum, finite temperature medium effects, although well studied for radiative decay~\cite{DOlivo:1989brs,Nieves:1997md,Grasso:1998td}, have received comparatively little attention. 

In this work, we have investigated neutrino decay in the presence of a thermal bosonic background using the framework of finite-temperature quantum field theory. Starting from the imaginary part of the neutrino self-energy, we derived the thermal decay rate for a heavier neutrino decaying into a lighter neutrino accompanied by a scalar or vector boson. The resulting expressions incorporate the effects of the thermal medium through the occupation numbers of the particles participating in the decay and naturally reproduce the corresponding vacuum results in the zero-temperature limit.

A central result of this work is the identification of a significant thermal enhancement of the decay rate in the quasi-degenerate regime. When the mass splitting between the parent and daughter neutrinos is small, the emitted boson becomes kinematically soft. For relativistic parent neutrinos, Lorentz boosting further suppresses the boson energy in the laboratory frame, leading to large bosonic occupation numbers. We have shown that this enhancement arises from a generic interplay between finite-temperature effects and quasi-degenerate kinematics and is largely independent of the specific nature of the bosonic mediator.

We applied the formalism to non-standard neutrino decays mediated by both scalar and vector interactions and demonstrated that if neutrinos are highly degenerate and propagate in a thermal bosonic background, the decay width can be increased by orders of magnitude compared to such a width in vacuum. Interestingly, this increase is independent of whether neutrinos decay into a scalar or a light gauge boson. Since many existing limits on neutrino lifetimes are derived assuming vacuum decay rates, our results suggest that medium-induced effects may play an important role whenever neutrinos propagate through environments containing an appreciable bosonic background.

As an illustrative application, we discuss the possible implications of the enhancement for cosmological neutrino decays. In the early Universe, thermal backgrounds are naturally present and may alter the competition between neutrino decay rates and the Hubble expansion rate. Consequently, the thermal enhancement identified here could modify existing constraints on neutrino decay or affect freeze-out scenarios with degenerate fermions, as in Ref.~\cite{Foguel:2024lca}. Since it is crucial for this effect to have the presence of the background, one could naively expect strong constraints on such a population of light bosons from Big-Bang Nucleosynthesis (BBN) or CMB from the measurement of the relativistic degrees of freedom, parametrized via $\Delta N_{\rm eff}$. Additionally, depending on the strength of the interaction, one could also induce modifications on the abundance of primordial elements. A possible way out is to consider models in which the scalar background reaches thermal equilibrium with neutrinos after BBN and decouples before CMB formation, as proposed in Ref.~\cite{Berbig:2020wve}. 

We emphasize that the formalism developed here is largely model-independent and relies only on the existence of a bosonic medium with a non-vanishing occupation number. While neutrino decay provides a particularly well-motivated application, the enhancement mechanism identified in this work is generic and may arise in a broad class of fermionic decay processes occurring in thermal environments. 
As a possible application, we can consider inelastic Dark Matter models, in which the Dark Matter is the lightest partner in an almost degenerate pseudo-Dirac pair, see e.g. Ref.~\cite{Foguel:2024lca}. We could expect that the thermal corrections we have computed could lead to modifications to the final relic density if a thermal background is allowed to be present during Dark Matter production. Additionally, this mechanism has been shown to be relevant for the production of light thermal bosonic DM, produced from decay of dark degenerate fermions~\cite{Yin:2023jjj,Sakurai:2024apm}.
This formalism might also have important implications for scenarios where the fermion mass and decay eigenstates do not overlap~\cite{Berryman:2014yoa, Chattopadhyay:2022isn, Abada:2023mib,Abada:2025gvc}. Furthermore, such stimulated emission/absorption of non-thermal bosons by neutrinos have also been used to interpret the phenomenon of collective neutrino oscillations in a core-collapse supernova~\cite{Fiorillo:2025npi}, as well as for studying neutrino decays inside the Sun~\cite{Malaney:1994qj}.

Another relevant model that could be affected by this thermal enhancement is leptogenesis~\cite{Fukugita:1986hr}. In this case, the scalar background would naturally be present as it corresponds to the Higgs sector. 
However, to have the effect, the masses of the additional right-handed neutrinos would need to be significantly close to the lepton (thermal) masses, so that this decay augmentation might only be present for a specific region of the parameter space. Given these difficulties, we leave for future work the specific application of the thermal enhancement derived here to a realistic model of neutrinos or Dark Matter.

In summary, our results demonstrate that finite-temperature effects can qualitatively modify neutrino decay rates in the presence of a bosonic background. The enhancement found in the quasi-degenerate regime represents a previously underappreciated aspect of neutrino decay phenomenology and motivates further studies of medium-induced effects in both cosmological and astrophysical environments.

\section*{Acknowledgements}

MS acknowledges support from the Early Career Research Grant by Anusandhan National Research Foundation (project number ANRF/ECRG/2024/000522/PMS). MS also acknowledges support from the IoE-funded Seed Funding for Collaboration and Partnership Projects - Phase IV SCPP grant (RD/0524-IOE00I0-012) by IIT Bombay. WT is supported by the National Science Foundation under Grant No.~PHY-2310224. WT is grateful to the Physics Department at Northwestern University and the Physics Department at IIT-Bombay for their hospitality during the initial stages of this work. We would also like to thank the Galileo Galilei Institute for Theoretical Physics for the hospitality and the INFN for partial support during the conception of this work.

\bibliographystyle{apsrev4-1}
\bibliography{referencesNuDecay}

\appendix

\section{Calculation of the neutrino decay in a thermal bath} \label{app:calculation}

We consider a neutrino in a thermal ensemble with temperature $T$. As presented in \cite{Weldon:1983jn}, the self-energy of the neutrino, $\Pi(\omega)$, accounts for the decay and inverse decay processes,
\begin{equation} \label{eq:decay-gamma-app}
\operatorname{Im} \Pi(\omega)=-\omega\,\Gamma(\omega),
\end{equation} where $\Gamma$ is the sum of the decay and inverse-decay rates, $\Gamma\,=\, \Gamma_D+\Gamma_I$. $\Gamma$ represents the rate at which the parent population relaxes toward equilibrium. Since the decay and inverse-decay are given by the same matrix element and the fermionic distribution functions obey $(1-f_{FD})/f_{FD}\,=\,e^{\omega/T}$, the rates are related by 
\begin{equation}
    \frac{\Gamma_D}{\Gamma_I}\,=\,e^{\omega/T}.
\end{equation}
As a consequence, the decay rate is written as
\begin{equation} \label{eq:decay-def-app}
\Gamma_D(\omega)\,=\,-\frac{1}{1+e^{-\omega/T}}  \frac{\operatorname{Im} \Pi(\omega)}{\omega}.
\end{equation}
Notice that $\Gamma$ and $\Gamma_D$, while having different meanings, differ only by the factor $1+e^{-E/T}$, so every enhancement shown in our work for the decay rate applies equally to the population relaxation rate.

To calculate the self-energy due to the neutrino's interaction with a scalar field, Eq.~\eqref{eq:Lagrangian-int}, we use the Matusbara frequencies, $p^0\,=\,i \omega_n\,=\,2n\beta \pi\,i $ for bosons, and $p^0\,=\,i \omega_n\,=(2n+ 1)\beta \pi\,i$. The amplitude of the diagram in Figure~\ref{fig:self-energy} is given by 
\begin{equation}
\label{eq:self-energy-app}
    -i \Sigma(p)\,=\,g^2 \int \frac{dp^0_1}{2\pi} \int \frac{d^3p_1}{(2\pi)^3}\,\mathrm{P_R }\,S(p_1)\mathrm{P_L }\,D(p-p_1), 
\end{equation}where $S(p)$ and $D(p)$ are the fermionic and scalar propagators respectively, and $P_{L,R}$ are the left and right projection operators. Implementing the Matsubara formalism, these propagators become~\cite{Bellac:2011kqa}
\begin{eqnarray}
 S(p_1)&=&\frac{\slashed{p}+m}{p_1^2-m^2}  \,=\,- \frac{i\omega_n\gamma^0 - \vec{p_1}\cdot \vec{\gamma} +m}{\omega_n^2+\vec{p}_1^{\,2}+m^2} \,\,=\,\, - \frac{i\omega_n\gamma^0 - \vec{p_1}\cdot \vec{\gamma} +m}{\omega_n^2+E_1^2},\\
D(p-p_1) &=& \frac{1}{(p-p_1)^2-m_\phi^2}\,=\,-\frac{1}{(\omega-\omega_n)^2+E_2^2}, \\
&& {\rm where}\quad E_1\equiv \sqrt{\vec{p}_1^{\,2}+m^2},\quad E_2\,\equiv\, \sqrt{(\vec{p}-\vec{p}_1)^{2}+m_\phi^2}. \nonumber
\end{eqnarray}

On the other hand, integration over $p_1^0$ becomes the sum  $\int \frac{dp^0_1}{2\pi} \rightarrow iT\sum_{n}. $
Thus, the self-energy becomes
\begin{eqnarray} \label{eq:self-ene-app-2}
\Sigma(p)&=&-\,g^2 \,T \sum_{n=-\infty}^{\infty} \int \frac{d^3p_1}{(2\pi)^3}\mathrm{P_R }\frac{i\omega_n\gamma^0 - \vec{p_1}\cdot \vec{\gamma} +m}{\omega_n^2+E_1^2}\mathrm{P_L } \frac{1}{(\omega-\omega_n)^2+E_2},\nonumber \\
&=& -\,g^2 \,T \sum_{n=-\infty}^{\infty} \int \frac{d^3\vec{p}_1}{(2\pi)^3}\, \frac{i\omega_n\gamma^0 - \vec{p_1}\cdot \vec{\gamma}}{\omega_n^2+E_1^2} P_L \frac{1}{(\omega-\omega_n)^2+E_2},
\end{eqnarray}

To carry out the sum in Eq.~\eqref{eq:self-ene-app-2}, we use the identities~\cite{Bellac:2011kqa}
\begin{footnotesize}
\begin{eqnarray}
    T\sum_{n}^{\infty} \frac{1}{\omega_n^2+E_1^2} \frac{1}{(\omega-\omega_n)^2+E_2^2} &=& \frac{1}{4E_1 E_2}\left\lbrace \left(\frac{1}{i\omega +E_1-E_2} - \frac{1}{i\omega -E_1+E_2}\right)\left(f_{FD}(E_1)+f_{BE}(E_2) \right) \right. \nonumber\\
    && + \left.\left(\frac{1}{i\omega +E_1+E_2} - \frac{1}{i\omega -E_1-E_2}\right)\left(1-f_{FD}(E_1)+f_{BE}(E_2) \right)  \right\rbrace,  \\
    &\equiv& F_1(i\omega, E_1,E_2) \nonumber\\
    T\sum_{n}^{\infty} \frac{\omega_n}{\omega_n^2+E_1^2} \frac{1}{(\omega-\omega_n)^2+E_2^2} &=& \frac{i}{4E_2}\left\lbrace \left(\frac{1}{i\omega +E_1-E_2} + \frac{1}{i\omega -E_1+E_2}\right)\left(f_{FD}(E_1)+f_{BE}(E_2) \right) \right. \nonumber\\
    && + \left.\left(\frac{1}{i\omega +E_1+E_2} + \frac{1}{i\omega -E_1-E_2}\right)\left(1-f_{FD}(E_1)+f_{BE}(E_2) \right)  \right\rbrace \\
        &\equiv& i\,F_2(i\omega, E_1,E_2) \nonumber
 \end{eqnarray}
\end{footnotesize}
where $f_{FD}$ and $f_{BE}$ are the Fermi-Dirac and Bose-Einstein distribution functions respectively.

So we can write Eq.~\eqref{eq:self-ene-app-2} as
\begin{equation} \label{eq:self-ene-3}
\Sigma(p)\,=-\,g^2  \int \frac{d^3p_1}{(2\pi)^3} (-\gamma^0 F_2(i\omega, E_1,E_2) - \vec{p_1}\cdot \vec{\gamma} \,F_1(i\omega, E_1,E_2))\mathrm{P_L }.
\end{equation}

The imaginary part of the self-energy can be obtained by calculating the discontinuity~\cite{Bellac:2011kqa}, 
\begin{equation} \label{eq:discontinuity-ap}
    \operatorname{Disc}\Sigma(\omega)\,=\,\,2i \operatorname{Im}\Sigma(\omega)\,=\,\Sigma\left(\omega+i\eta\right) -\Sigma\left(\omega-i\eta\right) .
\end{equation}
We redefine $i\omega\rightarrow p^0$ and use the discontinuity of $\Sigma(p)$ to obtain
\begin{eqnarray} \label{eq:ImPi_PL}
\operatorname{Im} \Sigma(p)\,=\, -\frac{g^2 \pi}{4} \int \frac{d^3 p_1}{(2\pi)^3 E_1E_2} &&\left\lbrace  \left(\gamma^0E_1+\vec{p_1}\cdot\vec{\gamma}\right)\,\delta\left(p^0+E_1-E_2\right)(f_{FD}(E_1)+f_{BE}(E_2))  \right. \\  
&& \left.  +\left(\gamma^0E_1-\vec{p_1}\cdot\vec{\gamma}\right)\,\delta\left(p^0-E_1+E_2\right)(f_{FD}(E_1)+f_{BE}(E_2)) \right. \nonumber\\
&& \left. +\left(\gamma^0E_1+\vec{p_1}\cdot\vec{\gamma}\right)\,\delta\left(p^0+E_1+E_2\right)(1-f_{FD}(E_1)+f_{BE}(E_2)) \right. \nonumber\\
&& \left. +\left(\gamma^0E_1-\vec{p_1}\cdot\vec{\gamma}\right)\,\delta\left(p^0-E_1-E_2\right)(1-f_{FD}(E_1)+f_{BE}(E_2)) \right \rbrace\,\mathrm{P_L } ,\nonumber
\end{eqnarray}
where we have used the identity
\begin{equation}
  \delta(x) \,=\,   \lim_{\eta \rightarrow 0} \frac{1}{\pi} \frac{\eta}{x^2 +\eta^2} \,.
\end{equation}

Notice that, in Eq.~\eqref{eq:ImPi_PL}, we can rewrite $1-f_{FD}(E_1)+f_{BE}(E_2)\,=\,(1-f_{FD}(E_1))\,(1+f_{BE}(E_2))+f_{FD}(E_1)f_{BE}(E_2)$, which can be identified with the inverse decay and decay, $\nu \rightleftharpoons \nu_{\rm bkg}\phi$ (third and fourth line of Eq.~\eqref{eq:ImPi_PL}, respectively).

Now, we define 
 \begin{eqnarray}
    \Pi(p) &=& \frac{1}{2}\sum_{\rm spin}\, \overline{u}(p) \,\Sigma (p) \,u(p) \,=\,\frac{1}{2} \operatorname{Tr}\left[(\slashed{p}+M) \Sigma(p)\right]\, ,
\end{eqnarray} and carry out the sum over spin to obtain

\begin{eqnarray}
    {\rm Im}\Pi(p) &=\, -\displaystyle\frac{g^2 \pi}{4} \int \displaystyle\frac{d^3 p_1}{(2\pi)^3 E_jE_\phi} &\left\lbrace p\cdot \tilde{p}_1 \left[\delta\left(p^0+E_1-E_2\right)  +\delta\left(p^0-E_1+E_2\right)\right](f_{FD}(E_1)+f_{BE}(E_2)) \right. \\
&& \left. +\,p\cdot p_1\left[\delta\left(p^0+E_1+E_2\right)+\delta\left(p^0-E_1-E_2\right)\right](1-f_{FD}(E_1)+f_{BE}(E_2)) \right \rbrace\, , \nonumber
\end{eqnarray}
where $p\cdot \tilde{p}_1\,=\,p^0E_1+\vec{p}\cdot \vec{p}_1.$ 

Focusing on the decay process, given by the term proportional to $\delta\left(p^0-E_1-E_2\right)$ in the equation above, 
\begin{align} \label{eq:ImPi}
 \operatorname{Im} \Pi(p)&= -\displaystyle\frac{g^2 \pi}{4} \int \frac{d^3p_1}{(2\pi)^3 \,E_1E_2} \left\lbrace  p\cdot p_1\,\delta\left(p^0-E_1-E_2\right)(1-f_{FD}(E_1)+f_{BE}(E_2))\right \rbrace , \\
  &= \, -\displaystyle\frac{g^2 \pi}{4}\frac{1}{(2\pi)^2} \int \frac{dE_1 |\vec{p}_1|}{E_2}\int d\cos\theta \left(p^0E_1-|\vec{p}|\,|\vec{p}_1|\,\cos\theta\right) \,\frac{E_2}{|\vec{p}|\,|\vec{p}_1|}\delta\left(\cos\theta-\cos\theta_*\right)\left(1-f_{FD}(E_1)+f_{BE}(E_2)\right),\nonumber\\
  &= \, -\displaystyle\frac{g^2 }{32\pi}\int \frac{dE_1 }{|\vec{p}|}\int d\cos\theta \left(p^0E_1-|\vec{p}|\,|\vec{p}_1|\,\cos\theta\right)  \,\delta\left(\cos\theta-\cos\theta_*\right)\left(1-f_{FD}(E_1)+f_{BE}(E_2)\right),\nonumber
\end{align}
where
\begin{equation}
     {\cos\theta_*\,\equiv\, \frac{2E_1p^0+m_\phi^2-M^2-m^2}{2\,|\vec{p}|\,|\vec{p}_1|}}.
\end{equation}
Integrating over $\cos\theta$ imposes that $\omega_{-}<E_1<\omega_{+}$, with
\begin{equation}
\omega_\pm \,=\, \frac{p^0}{2}\left(1-\frac{m_\phi^2-m^2}{M^2}\right)\pm \frac{p}{2}\sqrt{\left(1-\frac{(m_\phi-m)^2}{M^2}\right)\left(1-\frac{(m_\phi+m)^2}{M^2}\right)}.
\end{equation}

Integration over $E_1$, and redefining $E\equiv p^0$, results then in
\begin{align} 
 \operatorname{Im} \Pi(p) &=  \, - \frac{g^2}{32\pi\,p}(M^2+m^2-m_\phi^2)\, T  \log \left[\frac{\left(1+e^{\omega_{+}/T}\right)\left(e^{E/T}-e^{\omega_{-}/T}\right)}{\left(1+e^{\omega_{-}/T}\right)\left(e^{E/T}-e^{\omega_{+}/T}\right)} \right].
\end{align}

Finally, using Eq.~\eqref{eq:decay-def-app}, we write the decay rate as
\begin{eqnarray}
 \Gamma_D(E)&=&  \frac{g^2 (M^2+m^2-m_\phi^2)}{32\pi\,E\,\sqrt{E^2-M^2}}  \frac{T}{1+e^{-E/T}} \log \left[\frac{\left(1+e^{\omega_{+}/T}\right)\left(e^{E/T}-e^{\omega_{-}/T}\right)}{\left(1+e^{\omega_{-}/T}\right)\left(e^{E/T}-e^{\omega_{+}/T}\right)} \right].
\end{eqnarray} 

Where we can identify the vacuum decay rate,
\begin{equation}
 \Gamma_{\rm vac}(E) \,=\,  \frac{g^2 (M^2+m^2-m_\phi^2)}{32\pi\,E}  \sqrt{\left(1-\frac{(m-m_\phi)^2}{M^2}\right)\left(1-\frac{(m+m_\phi)^2}{M^2}\right)} \,. 
\end{equation} 

The terms in the first two lines of \eqref{eq:ImPi_PL}, proportional to $f_{\rm FD}+f_{\rm BE}$, correspond to the so-called ``Landau-damping'' associated to the processes $\nu_2\bar{\nu}_1\rightarrow \phi$ and $\nu_2\phi\rightarrow\nu_1$. These processes are kinematically closed for $M\,>\,m+m_\phi$ and therefore vanish throughout our analysis.

\section{Thermal decay of a fermion in the case of a vector mediator}
\label{app:vector}
In the case of a vector mediator, we replace the scalar propagator by the gauge boson propagator, in the unitary gauge,
\begin{equation}
 D_{\mu\nu}(q)\,=\, \frac{-\eta_{\mu\nu}+\frac{q_\mu q_\nu}{m_A^2}}{q^2-m_A^2}\,.
\end{equation}
We can now follow the same procedure as for a scalar. The only difference is the structure of the trace, where now we have 
\begin{eqnarray}
    \operatorname{Tr}\left[(\slashed{p}+M) \Sigma(p)\right] &\supset& \operatorname{Tr}\left[(\slashed{p}+M)\gamma^\mu (\slashed{p}_1+m)\gamma_\mu\,\mathrm{P_L }  - \frac{1}{m_A^2} (\slashed{p}+M)(\slashed{p}-\slashed{p}_1)(\slashed{p}_1+m)(\slashed{p}-\slashed{p}_1)\mathrm{P_L }\right] \nonumber  \\
    &=&\frac{1}{m_A^2} \left (4\,m^2M^2 -2\,p_1\cdot p\,(M^2+m^2+m_A^2) \right). 
\end{eqnarray}
The thermal decay rate is, after making $E\equiv p^0$, then
\begin{eqnarray}
 \Gamma_D(E)&=&  \frac{g^2}{16\pi E\,\sqrt{E^2-m_A^2}}  \left(M^2+m^2-2m_A^2+\frac{(M^2-m^2)^2}{m_A^2}\right) \frac{1}{1+e^{-E/T}}\,\log \left[\frac{\left(1+e^{\omega_{+}/T}\right)\left(e^{E/T}-e^{\omega_{-}/T}\right)}{\left(1+e^{\omega_{-}/T}\right)\left(e^{E/T}-e^{\omega_{+}/T}\right)} \right], 
\end{eqnarray} 
with
\begin{equation}
\quad\omega_{\pm} \,=\, \frac{E}{2}\left(1-\frac{m_A^2 - m^2}{M^2}\right)\pm \frac{\sqrt{E^2-M^2}}{2}\sqrt{\left(1-\frac{(m_A-m)^2}{M^2}\right)\left(1-\frac{(m_A+m)^2}{M^2}\right)}
\end{equation}


\section{Approximate Derivation of the Thermal Enhancement}
\label{app:enhancement}
In this Appendix, we summarize the origin of the enhancement appearing in the thermal decay rate in the quasi-degenerate limit in the scalar decay channel, $\nu_h(M) \rightarrow \nu_l(m) + \phi(m_\phi),$. We define the mass splitting as $\delta \equiv M - m$ and focus on the quasi-degenerate regime $\delta \ll M.$ In the rest frame of the parent neutrino, one obtains the standard two-body result
\begin{equation}
|\vec{p}_l| = |\vec{p}_\phi| \equiv p = \frac{1}{2 M} \sqrt{ \left[M^2-(m_l+m_\phi)^2\right]\left[M^2-(m_l-m_\phi)^2\right]}.
\end{equation}
In the limit $m_\phi \rightarrow 0$, this reduces to $p \simeq \delta $, therefore, in the quasi-degenerate regime the emitted scalar is kinematically soft in the parent rest frame.

Boosting to a frame where the parent neutrino moves along the z-axis with an energy $E$ and velocity $\beta$, the scalar energy becomes 
\begin{equation}
E_\phi^{\rm lab} =\gamma(E_\phi + \beta p\cos\theta),
\end{equation}
where $\gamma = E/M$ is the Lorentz factor and $\theta$ is the angle between the scalar momentum and the boost direction. Focussing on the specific case of backward emission $(\cos\theta = -1)$, we find 
\begin{equation}
E_\phi^{\rm lab} \simeq \frac{p}{2\gamma}\simeq\frac{\delta}{2\gamma}.
\end{equation}
For highly boosted parent neutrinos, $\gamma \gg 1$, the scalar energy becomes strongly suppressed.

The thermal correction to the decay rate contains the statistical factor $(1-f_{FD}+f_{BE})$, where, in the limit $E_\phi \ll T$, the Bose-Einstein distribution reduces to
\begin{equation}
 f_{BE}(E_\phi)\simeq\frac{T}{E_\phi}.   
\end{equation}
Substituting the boosted scalar energy in the lab frame gives
\begin{equation}
 f_{BE}\simeq \frac{2\gamma T}{\delta}.
\end{equation}
Thus, for quasi-degenerate neutrinos and relativistic parent energies, the bosonic occupation number can become parametrically large.

In this limit, the expression for the thermal decay rate, $\Gamma_D(E)$ in Eq.\,(\ref{eq:GammD}), simplifies drastically. In particular, Eq.\,(\ref{eq:wpm}) can be approximated as
\begin{equation}
\omega_\pm \simeq E \pm \frac{p\delta}{M}.
\end{equation}
Expanding the exponential factors in Eq.\,(\ref{eq:GammD}) for small $\delta/T$ yields an approximate logarithmic contribution of the form $e^{E/T}\left(p\,\delta\right)/\left(M\, T\right)$. Consequently, we get
\begin{equation}
\frac{\Gamma_{D}}{\Gamma_{\rm vac}} \simeq \frac{1}{2}e^{E/T}\,.
\end{equation}
The enhancement, therefore, becomes significant when the parent neutrino is 
relativistic and the emitted scalar is sufficiently soft.

\medskip

\end{document}